\documentclass[sagev, times, twocolumn]{sagej}

\usepackage{moreverb,url}
\usepackage{setspace}
\usepackage{bm}
\usepackage{multirow}
\usepackage{makecell}

\usepackage[colorlinks,bookmarksopen,bookmarksnumbered,citecolor=red,urlcolor=red]{hyperref}
\usepackage{draftwatermark} 
\SetWatermarkLightness{ 0.9 } 
\SetWatermarkText{DRAFT} 
\SetWatermarkScale{ 0.5}

\newcommand\BibTeX{{\rmfamily B\kern-.05em \textsc{i\kern-.025em b}\kern-.08em
T\kern-.1667em\lower.7ex\hbox{E}\kern-.125emX}}

\def\volumeyear{2022}

\runninghead{Hosseini et al.}

\title{Designing a Bayesian adaptive clinical trial to evaluate novel mechanical ventilation strategies in acute respiratory failure using Integrated Nested Laplace Approximations}

\author{Reyhaneh Hosseini\affilnum{1}, Ziming Chen\affilnum{1}, Ewan Goligher\affilnum{2}, Eddy Fan\affilnum{2, 6, 7}, Niall D. Ferguson\affilnum{2, 6, 7},  Michael O. Harhay\affilnum{3}, Sarina Sahetya\affilnum{4}, Martin Urner\affilnum{5,6}, Christopher J. Yarnell\affilnum{2,7} and Anna Heath\affilnum{1, 8, 9}}

\affiliation{
\affilnum{1} Child Health Evaluative Sciences, Peter Gilgan Centre for Research and Learning, The Hospital for Sick Children, Toronto, ON, Canada\\
\affilnum{2} Department of Medicine, Division of Respirology, University Health Network, Toronto, ON, Canada \\
\affilnum{3} Perelman School of Medicine, University of Pennsylvania, Philadelphia, PA, US \\
\affilnum{4} Division of Pulmonary and Critical Care Medicine, Johns Hopkins Hospital, Baltimore, MD, United States \\
\affilnum{5} Department of Anesthesiology and Pain Medicine, University of Toronto, Toronto, Canada\\
\affilnum{6} Interdepartmental Division of Critical Care Medicine, University of Toronto, Toronto, Canada \\
\affilnum{7} Insititute of Health Policy, Management, and Evaluation, University of Toronto, Toronto, Canada \\
\affilnum{8} Division of Biostatistics, Dalla Lana School of Public Health, University of Toronto, Toronto, ON, Canada \\
\affilnum{9} Department of Statistical Science, University College London, London, UK
}

\corrauth{Anna Heath, Child Health Evaluative Sciences, Peter Gilgan Centre for Research and Learning, The Hospital for Sick Children, 686 Bay Street, Toronto,
ON M5G 0A4, Canada.
Email: anna.heath@sickkids.ca}

\email{anna.heath@sickkids.ca}

\begin{document}

\maketitle

\section*{Abstract}

\textbf{Background:} Trials with pre-planned adaptive design elements usually require extensive simulations to determine appropriate values for key design parameters, demonstrate error rates, and establish the required/expected sample size. This study performed simulations to design a Bayesian adaptive trial comparing ventilation strategies for patients with acute hypoxemic respiratory failure. However, the complexity of the proposed outcome and analysis meant that Markov Chain Monte Carlo methods would usually be required to estimate the posterior distributions during the simulations, requiring an infeasible computational burden. Thus, we have also leveraged the Integrated Nested Laplace Approximations (INLA) algorithm, a fast approximation method for Bayesian inference, to ensure the feasibility of these simulations.

\textbf{Methods:} We simulated Bayesian adaptive two-arm superiority trials with equal randomisation that stratified participants into two disease severity states. The outcome was the number of days alive and free of mechanical ventilation to day 28 and was analyzed with proportional odds logistic regression. Trials were stopped based on prespecified posterior probability thresholds for superiority or futility, separately for each state. We calculated the type I error and power across 64 scenarios that varied the posterior probability thresholds and the initial minimum sample size before commencing adaptive analyses. We incorporated dynamic borrowing in the models and used INLA to compute the posterior distributions at each adaptive analysis. Designs that maintained a type I error below 5\%, a power above 80\%, and a feasible mean sample size were then evaluated across 22 scenarios that varied the odds ratios for the two severity states to determine the optimal design for the trial.

\textbf{Results:} Power generally increased as the initial sample size and the threshold for declaring futility increased. Two designs maintained high power, a type I error below 5\% and a feasible mean sample size across both severity states. These designs had an initial recruitment of 500 and 200 and a threshold for declaring superiority of 0.9925 and 0.995, respectively, and a threshold for declaring futility of 0.95 for both. In the comprehensive simulations, the first design had a higher chance of reaching a trial conclusion before the maximum sample size and higher probability of declaring superiority when appropriate without a substantial increase in sample size for the more realistic scenarios, compared to second design, and the first design was chosen as the trial design. 

\textbf{Conclusions:} We designed a Bayesian adaptive trial to evaluate novel strategies for ventilation. The INLA algorithm allowed us to evaluate a wide range of designs through simulation and optimize the trial design by balancing the mean sample size and power.

\paragraph{Keywords} Bayesian clinical trial design, proportional odds model, acute hypoxemic respiratory failure, critical care medicine, Integrated Nested Laplace Approximations

\doublespacing

\section{Introduction}
Adaptive trials allow for pre-planned adjustments to trial conduct based on accumulating data in the trial, while maintaining the validity and integrity of the trial.\cite{dimairo2020adaptive, pallmann2018adaptive} A large range of adaptions can be proposed, and typically aim to improve trial efficiency, e.g., through reducing the required sample size, which saves time, resources and money.\cite{bauer2016twenty,pallmann2018adaptive} However, the operating characteristics of trial designs must be comprehensively evaluated and, with adaptive elements, this evaluation requires computationally challenging simulations to ensure the statistical validity of the proposed adaptions and subsequent analyses.\cite{thorlund2018key}

Bayesian methods for the design and analysis of adaptive trials are becoming increasingly popular\cite{biswas2009bayesian, tidwell2019bayesian, ashby2006bayesian} and can improve performance compared to frequentist adaptive trials.\cite{ryan2019using} Furthermore, Bayesian adaptive trials were used frequently during the COVID-19 pandemic to efficiently determine effective treatments.\cite{remap2021therapeutic, attacc2021therapeutic, ali2022remdesivir, reis2021effect} The Bayesian approach is particularly suited for adaptive trials as it supports progressive learning from accumulating data.\cite{lee1989bayesian} The Bayesian approach can also improve statistical efficiency using hierarchical modeling to combine information across different subgroups\cite{berry2010bayesian} and can synthesize information across multiple sources through the inclusion of prior information.\cite{lee1989bayesian}

The simulations required to design Bayesian adaptive trials are computationally expensive or even prohibitive if Markov Chain Monte Carlo (MCMC) methods are used to simulate from the posterior distribution.\cite{hansen2018practical, jack2012bayesian, chevret2012bayesian} Thus, the computational burden of these simulations is often simplified using conjugate distributions.\cite{berry2010bayesian} However, some clinical trial outcomes cannot be modelled using conjugate distributions, e.g., survival outcomes, which substantially limits the number of simulations that can be completed and therefore the complexity of the design. 

We aimed to develop a Bayesian adaptive trial design to evaluate a novel ventilation strategy for patients admitted into the intensive care unit (ICU) with acute hypoxemic respiratory failure (AHRF). In ICU patients, AHRF is a common and potentially life-threatening condition characterized by varying degrees of lung failure, for which patients often require significant respiratory support to maintain their gas exchange.\cite{cummings2020epidemiology, bernard1994american} The risk of death from AHRF varies between 20\% to 50\% \cite{bellani2016epidemiology} and survivors who receive significant respiratory support over a prolonged period of time often experience significant long-term disability.\cite{herridge2016recover} Interventions for AHRF therefore aim to reduce both mortality and the duration of time spent on the ventilator. The number of ventilator-free days is a common composite endpoint reflecting both of these patient-important effects and can be analysed as an ordinal outcome.\cite{attacc2021therapeutic,remap2021therapeutic} The use of an ordinal outcome means that conjugate distributions were not available to design this trial.

Integrated Nested Laplace Approximations (INLA) is an algorithm for efficient approximate Bayesian analysis for a broad class of models.\cite{rue2009approximate} This class of models includes the proposed ordinal model and can account for potential treatment effect heterogeneity across two subgroups, defined by the severity of AHRF. Using INLA to estimate the relevant posterior distributions, we evaluated 64 different trial designs to determine designs with type 1 error below 5\%, power above 80\% and a feasible sample size. We proposed frequent interim analyses and based trial conclusions on posterior probabilities of superiority and futility of the intervention compared to standard care. Our simulation study evaluated different thresholds for decision-making within the trial and the initial number of patients recruited before analysing the data. Finally, we comprehensively evaluated the best designs across 22 different assumptions about the treatment effect in each of the two severity groups. The simulation results showed that our novel Bayesian design will likely lead to a feasible sample size, while maintaining sufficient statistical power and valid inference.

This article begins by introducing the proposed trial design, including the primary outcome and proposed analysis model. We then introduce the INLA algorithm and the adjustments that were required to use INLA for the trial simulations. We also present the simulations used to comprehensively evaluate our design before summarising the results of the simulation study and determining the design of our trial.

\section{Methods}
\subsection{The DRIVE Trial}
The Platform of Randomized Adaptive Clinical Trials in Critical Illness (PRACTICAL; \url{https://practicalplatform.org/}) is a Bayesian adaptive platform randomized clinical trial studying novel interventions to improve outcomes for patients with AHRF. Within PRACTICAL, the Driving Pressure-Limited Ventilation in Hypoxemic Respiratory Failure (DRIVE) domain aims to compare the current standard care in ventilation (Usual Care; UC), which primarily aims to limit the volume of air inflating the lungs with each breath to avoid ventilator-induced lung injury,\cite{slutsky2013ventilator} to a novel ventilation strategy that limits driving pressure (Driving Pressure Limited; DPL), a cyclic pressure applied by the ventilator breath-by-breath, in addition to controlling maximum pressures in the alveoli.\cite{amato2015driving} Furthermore, as there is substantial heterogeneity in the severity of AHRF,\cite{bellani2016epidemiology} we will allow for potential heterogeneity in the treatment effect for patients in different severity states, defined by respiratory system elastance (high, $ \geq 2.5$ cm $H_2O/(mL/kg)$, or low, $<2.5$ cm $H_2O/(mL/kg)$). 

The DRIVE trial will use frequent analyses throughout the trial, planned every three months after reaching an initial enrolment requirement, to determine whether sufficient evidence has been collected to either conclude superiority or futility of DPL compared to UC, defined below. The goal of this design is to make conclusions as soon as possible by evaluating the therapies in repeated data analyses until one of these two trial termination triggers are met. The proposed design accounts for heterogeneity in the treatment effect according to elastance state (low and high) by making statistical conclusions of superiority and futility separately for each elastance state. Note that while DRIVE is embedded within the PRACTICAL platform trial, we do not currently expect interactions with other interventions in the platform and, thus, they are not be considered in this simulation study. 

\subsection{Primary Outcome and Data Generation}
To account for differences in mortality and length of time on the ventilator, the primary outcome in the DRIVE trial is a composite endpoint that combines in-hospital mortality and days alive and free from mechanical ventilation, known as ventilator-free days (VFDs).\cite{remap2021therapeutic, attacc2021therapeutic} We use ventilator-free days (VFDs) to day 28, defined as a 30-level ordinal categorical outcome. Death in hospital before day 90 is assigned a value of –1 with patients who remain alive and in hospital at day 90 considered as survivors. Survivors who spend more than 28 days on invasive ventilation are assigned a value of 0. All other outcomes between 1 and 28 are computed as the number of days alive and free of invasive ventilation in survivors who are liberated before day 28 after randomization. Once a patient is discharged from hospital, they are assumed to be alive and free of ventilation through to day 28. In this way, VFDs captures the effect of the intervention on survival and on time to liberation from mechanical ventilation. For the simulation study, VFDs can be simulated from a multinomial distribution $Mult(30,\boldsymbol{p})$, $\boldsymbol{p}=(p_{-1},\ldots,p_{28})$, where the relevant probabilities are estimated from available data, for UC, and generated from the assumed proportional odds model, for DPL. 

\subsection{Statistical Analysis and Modeling} 
We use Bayesian proportional odds logistic regression to model VFDs, which estimates the effect of DPL on VFDs as an odds ratio $OR$. An $OR < 1$ signifies benefit as DPL results in higher values for VFDs, i.e., the number of patients with an outcome of $-1$ reduces and patients are likely to experience fewer days on the ventilator. Conversely, $OR > 1$ indicates harm as DPL reduces VFDs. The DRIVE trial estimates separate odds ratios for patients in the high and low elastance states to account for potential heterogeneity in treatment effect.

To define the proportional odds model, let $\pi_j^i = P(Y \leq j)$ be the probability that the VFDs are less than or equal to $j$, $j = -1,\ldots 27$ for patient $i$, where $\pi_{28}^i = 1$ for all patients. Let $T_i$ be a treatment indicator, such that $T_i = 0$ if patient $i$ is randomised to UC and $T_i = 1$ if patient $i$ is randomised to DPL. Similarly, let $S_i$ be the elastance state for patient $i$, $S_i = 0$ for high elastance and $S_i = 1$ for low elastance. Bayesian dynamic borrowing can incorporate information from both elastance state subgroups to provide a more informative estimate of the treatment effect and achieve a more effective trial design by reducing sample size when appropriate. This is achieved through a hierarchical model that considers different treatment effects for each elastance states but assumes they come from the same distribution. With only two elastance states, the prior for the hierarchical model variance will have substantial influence on the final borrowing. Thus, we provide rationale for our prior selection in the supplementary material.

The outcome is modelled as \[ \log\left(\frac{\pi^{ik}_j}{1 - \pi^{ik}_j}\right) = \alpha_{j} - \beta_s S_{ik} - \beta_t T_{ik} - U_{k} T_{{ik}},\] where $j = -1,\ldots, 28,$, $\beta_s$ is the log odds ratio for patients receiving the DPL treatment being in high elastance group compared to low elastance group. $\beta_t$ is the log odds ratio for patients in the same elastance state receiving UC compared to DPL without accounting for variation between elastance states. $U_k$ differentiates the treatment effects for the two elastance states with $k = 1, 2$ for low and high elastance group, respectively, and is assumed to follow a normal distribution $N(0, \sigma^2)$. Thus, the log-odds ratio for patients with low elastance is $-(\beta_t + U_1)$ and $-(\beta_t + U_2)$ for patients with high elastance, comparing patients who received DPL to UC. We use minimally informative normal priors centered on 0 with a variance of 1000 for $\beta_s$ and $\beta_t$. A half-t prior with 3 degrees of freedom and a scale parameter of 7 is used on the hierarchical model variance $\sigma^2$ to account for relatively weak borrowing. The prior for the intercepts in the proportional odds model $\alpha_k$, $k = -1, \dots, 27$, is discussed below. 

\subsection{Stopping Rules for the Adaptive Design}
Our adaptive trial implements a sequence of analyses conducted every 190 patients, which is our expected recruitment rate in three months based on previous data. We fix that a third of patients (64) will have high elastance and two thirds of patients will have low elastance (126). At each analysis, we will fit the model outlined above using all available data to determine the posterior distribution of the log-odds ratio for both elastance states. Based on these posterior distributions, we will then evaluate whether either superiority or futility can be concluded for either of the elastance states. The two statistical triggers, superiority and futility, are defined as:
\begin{itemize}
\item The trial could stop for \textbf{superiority} of DPL compared to UC if the probability that the odds ratio of DPL compared to UC is less than 1 is higher than some threshold $p_{sup}$, i.e., superiority will be declared in the low elastance state if $ P(\text{exp}(-(\beta_t + U_1)) < 1) \geq p_{sup}$ and in the high elastance state if $ P(\text{exp}(-((\beta_t + U_2)) < 1) \geq p_{sup}$.
\item The trial could stop for \textbf{futility} of DPL compared to UC if the probability that DPL provides a less than 17\% improvement compared to DPL, as measured by the odds ratio of $\frac{1}{1.2}$, is greater than $p_{futi}$, i.e., futility will be declared in the low elastance state if $ P(\text{exp}(-((\beta_t + U_1))) > \frac{1}{1.2}) \geq p_{futi}$ and in the high elastance state if $ P(\text{exp}(-(\beta_t + U_2)) > \frac{1}{1.2}) \geq p_{futi}$.
\end{itemize}

The futility trigger stops the trial if there is mounting evidence that the magnitude of benefit is not sufficient to justify continuing the trial. With an odds ratio of 1.2 comparing UC to DPL, the required 17\% improvement was determined as the minimally important clinical difference for VFDs and represents an approximate 4\% absolute reduction in mortality based on the baseline event rates seen in previous data (supplementary material). The DRIVE trial is designed to continue enrolling indefinitely until a statistical trigger has been observed in both elastance states. However, to ensure feasibility of the simulation procedure and mimic the setting where all trials are completed, we set the maximum sample size at 5,000 patients per elastance state. Practically, this maximum sample size represents an infeasible trial size for DRIVE, based on previous trials in this patient population \cite{cavalcanti2017effect}. Finally, to guard against incorrect results due to small sample sizes at the early interim analyses, the first interim analysis for each state will not take place until $n_{init}$ patients have been enrolled.

\section{Simulations for the DRIVE Trial}
Simulations were used to inform the DRIVE trial design. The first simulations aim to find a set of valid and efficient designs for the DRIVE trial by varying the design parameters, $p_{sup}$, $p_{futi}$ and $n_{init}$ for two odds ratios. The second simulations then aimed to evaluate the chosen designs across different values for the expected treatment effect. 

\subsection{Integrated Nested Laplace Approximations}
To reduce the computational complexity of the proposed trial simulations, we used INLA,\cite{rue2017bayesian} an algorithm for efficient, approximate Bayesian analysis, to determine the posterior distribution of the log-odds ratios. INLA computes an analytic approximation for the posterior log-odds ratio by noting that a proportional odds models can be expressed as a Gaussian Markov Random Field,\cite{rue2009approximate} a flexible class of models that requires specific hierarchical relationships between parameters to enable efficient computation. In fact, the range of models that can be represented in this form include hierarchical generalised linear regression models, survival models (e.g., cox proportional hazard models) and spatio-temporal models, meaning that INLA can facilitate simulations for the majority of proposed clinical trial outcomes. 

However, to maintain its computational efficiency, INLA requires a relatively small number of categories for ordinal outcomes. This means that the \texttt{R-INLA} package for implementing INLA restricts the number of categories for ordinal outcomes to fewer than 10 categories.\cite{martins2013bayesian} Thus, we re-categorized the 30 level VFDs outcome to a 9 level ordinal outcome, ensuring that the category representing death (-1) was maintained as a separate category. Theoretically, proportional odds ratios do not change if we move, combine, or delete categories\cite{stromberg1996collapsing}, meaning that we can reduce the number of categories and take advantage of the computational efficiency of the INLA algorithm without invalidating the simulation results. As ordinal data have maximum power when the probabilities of being within each category are even,\cite{taylor2006loss} our collapsed ordinal outcome was created by combining categories to make the probability of each category as even as possible.

This means that the proportional odds model for our simulation study is \[ \log\left(\frac{\pi^{ik}_j}{1 - \pi^{ik}_j}\right) = \alpha_{j} - \beta_s S_{ik} - \beta_t T_{ik} - U_{k} T_{{ik}},\] with $j = 1,\ldots, 9$. The prior for $\alpha_j$ is specified by re-parameterising such that $\theta_1 = \alpha_1$ and $\alpha_k = \alpha_{k - 1} + e^{\theta_k},$ and using a Dirichlet prior with parameters 100 for $\bm\theta = (\theta_1, \dots, \theta_9)$. This prior was selected to ensure tractability of the simulations as numerical issues were sometimes encountered with small numbers of observations in some categories when a smaller value was used in the Dirichlet prior. 

\subsection{Generating the Ventilator Free Days Data}
The baseline event risk for each VFDs category was estimated from patients from the Toronto Intensive Care Observational Registry (iCORE) admitted to 9 ICUs from 7 different hospitals across the Greater Toronto Area. Data were extracted on 4\textsuperscript{th} Feburary 2022 and included 700 patients with low elastance and 160 with high elastance enrolled between 2014 and 2022. Table \ref{tab:baseline} displays the proportion of patients in the nine ordinal VFDs categories with the current standard care (highlighted in bold). Patients with high elastance have higher chances of experiencing lower VFDs, compared to the low elastance state, highlighting the heterogeneity in outcomes.

\begin{table*}[h]
\caption{The probability of experiencing each outcome for the nine-level ordinal outcome used in the simulations for the DRIVE trial. These probabilities are provided separately for the low and high elastance states and for a range of assumed values for the odds ratio, 1/0.8, 1, 1/1.1, 1/1.2, 1/1.25, 1/1.3, and 1/1.5. The correspondence between the collapsed ordinal outcome and the original 30-level VFDs outcome is also provided.}
\begin{center}
\begin{small}
\begin{tabular}{cccccccccc}
    \toprule
    \multirow{2}{*}{Elastance} & \multirow{2}{*}{VFDs} & \multirow{2}{*}{VFDs Group} &
      \multicolumn{7}{c}{Odds Ratio} \\
         \cline{4-10} 
     &&& 1/0.8 & \textbf{1} & 1/1.1 & 1/1.2 & 1/1.25 & 1/1.3 & 1/1.5 \\
      \midrule
   \multirow{9}{*}{Low}  & -1 & 1 & 0.33 & \textbf{0.28} & 0.26 & 0.24 & 0.24 & 0.23 & 0.21\\
    & 0-7 & 2 & 0.07 & \textbf{0.07} & 0.07 & 0.06 & 0.06 & 0.06 & 0.06 \\
    & 8-14 & 3 & 0.07 & \textbf{0.07} & 0.07 & 0.07 & 0.07 & 0.06 & 0.06\\
    & 15-19 & 4 & 0.09 & \textbf{0.09} & 0.09 & 0.09 & 0.9 & 0.9 & 0.9\\
    & 20-22 & 5 & 0.1 & \textbf{0.1} & 0.1 & 0.1 & 0.1 & 0.1 & 0.1\\
    & 23-24 & 6 & 0.07 & \textbf{0.08} & 0.08 & 0.08 & 0.08 & 0.09 & 0.09\\
    & 25-26 & 7 & 0.11 & \textbf{0.12} & 0.13 & 0.14 & 0.14 & 0.14 & 0.15 \\
    & 27 & 8 & 0.06 & \textbf{0.07} & 0.07 & 0.08 & 0.08 & 0.08 & 0.09\\
    & 28 & 9 & 0.1 & \textbf{0.12} & 0.13 & 0.14 & 0.15 & 0.15 & 0.17\\
    \hline
   \multirow{9}{*}{High} & -1 & 1 & 0.46 & \textbf{0.4} & 0.38 & 0.36 & 0.35 & 0.34 & 0.31 \\
    & 0-7 & 2 & 0.11 & \textbf{0.11} & 0.11 & 0.1 & 0.1 & 0.1 & 0.1 \\
    & 8-14 & 3 & 0.06 & \textbf{0.06} & 0.06 & 0.06 & 0.06 & 0.06 & 0.06 \\
    & 15-19 & 4 & 0.08 & \textbf{0.09} & 0.09 & 0.09 & 0.09 & 0.09 & 0.09 \\ 
    & 20-22 & 5 & 0.06 & \textbf{0.07} & 0.07 & 0.07 & 0.07 & 0.08 & 0.08 \\
    & 23-24 & 6 & 0.06 & \textbf{0.07} & 0.07 & 0.08 & 0.08 & 0.08 & 0.08 \\
    & 25-26 & 7 & 0.09 & \textbf{0.11} & 0.11 & 0.12 & 0.12 & 0.13 & 0.14 \\
    & 27 & 8  & 0.04 & \textbf{0.04} & 0.05 & 0.05 & 0.05 & 0.05 & 0.06\\
    & 28 & 9 & 0.05 & \textbf{0.06} & 0.06 & 0.07 & 0.07 & 0.07 & 0.08\\
    \bottomrule
\end{tabular}
\end{small}
\end{center}
\label{tab:baseline}
\end{table*}

Based on the baseline event risk, we consider seven scenarios for the odds ratio of DPL compared to UC, $0.8$, which represents harm, $1$, which represent the null effect and $1.1, 1.2, 1.25, 1.3, 1.5$, which represent varying degrees of benefit. The probability of each ordinal outcome under these seven scenarios are displayed in Table \ref{tab:baseline}, separated by elastance state. Table \ref{tab:baseline} also provides the correspondence between the original VFDs categories and our nine ordinal categories. Finally, to facilitate the interpretation of the VFDs outcome, we provide the median, mean, inter-quartile range and probability of death for the VFDs for each of these scenarios in the supplementary material.

\subsection{Simulation Study Design}
The first simulation varied the design parameters $p_{sup}$, $p_{futi}$ and $n_{init}$ to determine suitable designs for the trial. We considered 4 levels for $p_{sup}$, the threshold for stopping the trial due to superiority of DPL, 0.98, 0.99, 0.9925, 0.995, 4 levels for $p_{futi}$, the threshold for stopping the trial due to futility for DPL, 0.85, 0.9, 0.95, 0.9925 and 4 levels for $n_{init}$, the number of participants recruited before starting the interim analyses, 200, 350, 500, 800. We evaluated these values in a fully factorial design by computing the type I error for each design, with odds ratio equal to 1, and the power, with odds ratio equal to 1.3. This requires a total of $128$ simulation scenarios with 1000 simulated trials for each scenario. 


For each simulation, the reason that the trial stopped, i.e., futility, superiority, or reaching the maximum sample size of 5,000, was recorded for each elastance state, alongside the number of patients recruited when the trial stopped. From these results, type I error was estimated as the proportion of trials that concluded superiority when the odds ratio was 1 and power is the proportion of trials that concluded superiority when the odds ratio was 1.3. We also computed the average number of recruited patients across all 1000 simulations for each scenario and elastance state. From these values, we extracted study designs with a type I error less than 5\% and a power above 80\% in both elastance states. From these designs, identified by their values for $p_{sup}$, $p_{futi}$ and $n_{init}$, those with feasible mean sample sizes were evaluated in the second simulation study. 

The second simulation study evaluated our chosen designs across 22 different assumptions for the odds ratio of DPL compared to UC across the two elastance states. We restricted the simulations to a feasible recruitment level of 2,000 patients per elastance state to evaluate the proportion of trials that are unable to reach a conclusion.\footnote{Note that the simulation was stopped at the first interim analysis after 2,000 patients had been recruited, leading maximum sample sizes that are just above 2,000.} 

We considered scenarios where the odds ratios were the same for the two elastance states and equal to 0.8, 1, 1.1, 1.2, 1.25, 1.3 and 1.5, as presented in Table \ref{tab:baseline}, scenarios where one state had a null treatment effect and the other state either caused harm or benefit and scenarios where states had the following odds ratios (1.1, 1.3) and (1.2, 1.3). We computed the proportion of studies for each outcome (futility, superiority, and no trigger met) and the 50\textsuperscript{th} and 80\textsuperscript{th} percentiles for the number of patients recruited across 1000 simulated trials for each scenario. These results were then used to select the final design by balancing power and expected recruitment levels to maximise the chances that the DRIVE trial will complete within a reasonable timeframe.

\section{Results}

\subsection{Selecting the Optimal Design Parameters}
\begin{figure*}[h]
\setlength{\fboxsep}{0pt}%
\setlength{\fboxrule}{0pt}%
\begin{center}
\includegraphics[width =\textwidth]{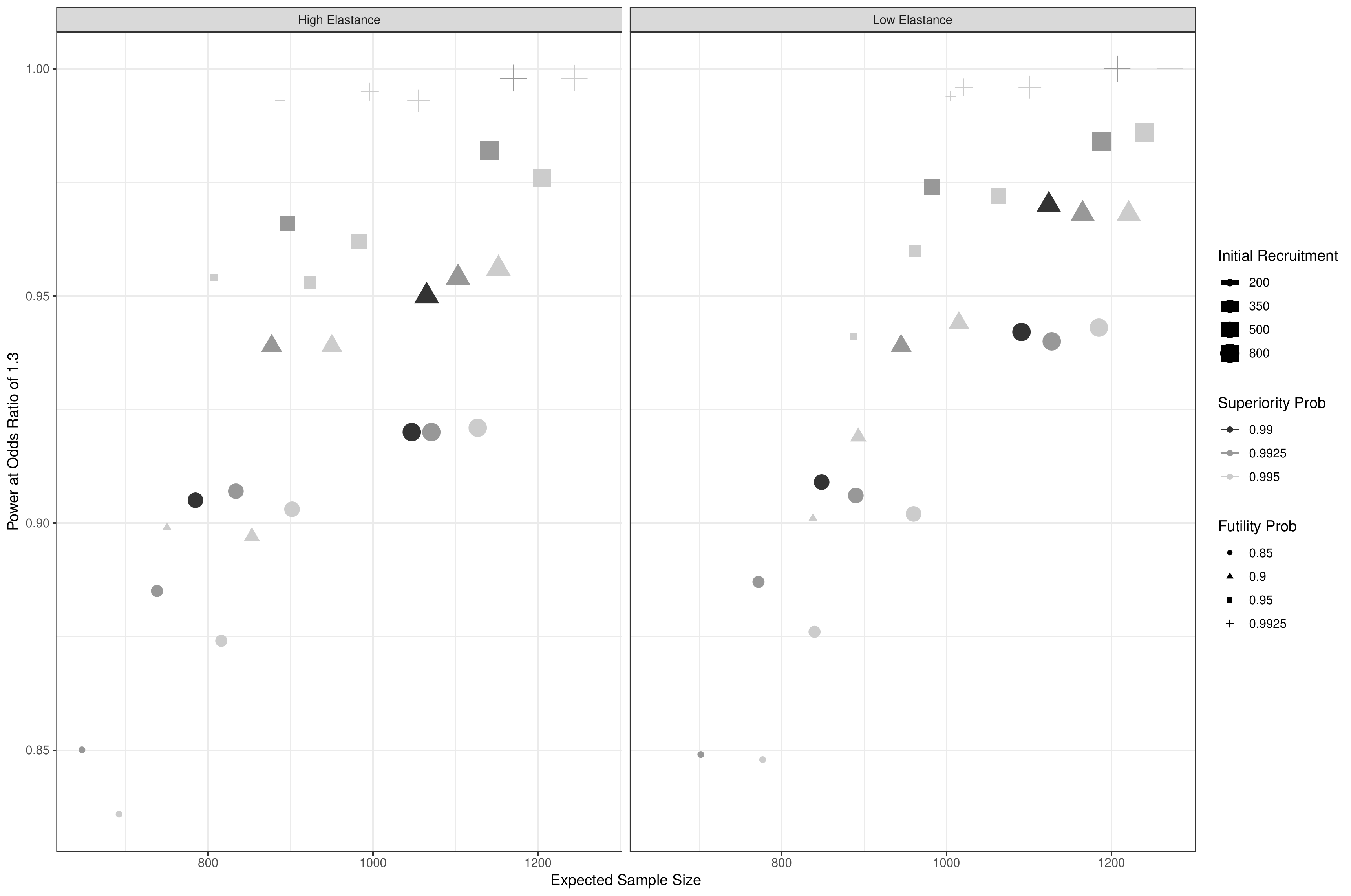}
\end{center}
\caption{The mean sample size and power calculated for the scenarios that maintained a type I error less than 5\% (selected from the $4\times4\times4=64$ scenarios considered) across different assumptions for the initial recruitment (350, 500, 800), probability threshold to stop for futility (0.85, 0.9, 0.95, 0.9925), probability threshold to stop for superiority (0.99, 0.9925, 0.995) for the two respiratory system elastance groups.\label{Fig:all_scen}}
\end{figure*}

Figure \ref{Fig:all_scen} displays the mean sample size and power for the designs with a type I error less than 5\% across both elastance states. Overall, a higher initial recruitment phase, represented by larger points in Figure \ref{Fig:all_scen}, resulted in a larger mean sample size and higher power. Lower values for $p_{futi}$, represented as different shapes in Figure \ref{Fig:all_scen}, resulted in lower mean sample sizes and a reduction in power. Note that as $p_{futi}$ increased, it was harder to control type I error, resulting in fewer scenarios plotted for higher values of $p_{futi}$. Higher values of $p_{sup}$, represented by different colours in Figure \ref{Fig:all_scen}, resulted in larger mean sample sizes, for a fixed initial recruitment level and the relationship with power was non-linear. If 0.98 is set as the posterior probability superiority threshold, the type I error was above 5\% for all scenarios and this value for $p_{sup}$ was not included in Figure \ref{Fig:all_scen}.

Figure \ref{Fig:all_scen} displays a general increasing trend in power as the mean sample size increases but some designs offer similar levels of power for smaller mean sample sizes. These designs lie on the left-hand edge of Figure \ref{Fig:all_scen} as this represents the design that provides the smallest sample size for a fixed value of power. From Figure \ref{Fig:all_scen}, we determined that the two designs displayed in Table \ref{tab:designs} offer the best balance between power and a feasible mean sample size.
\begin{table*}[!h]
\caption{The settings for 2 chosen designs from the first phase of simulation.}
\begin{center}
\begin{tabular}{|c|c|c|c|c|c|c|c|}
    \hline
    Design &  $n_{init}$ & $p_{sup}$ & $p_{fut}$ & State & Power & Type 1 Error & Mean Sample Size \\
    \hline
    \multirow{2}{*}{Design 1} & \multirow{2}{*}{500} & \multirow{2}{*}{0.9925} & \multirow{2}{*}{0.95} & LE & 0.974 & 0.040  & 982 \\ 
                        &  & & & HE & 0.966 & 0.042 & 896 \\  
    \hline
    \multirow{2}{*}{Design 2} & \multirow{2}{*}{200} & \multirow{2}{*}{0.995} & \multirow{2}{*}{0.95} & LE & 0.941 & 0.044  & 887 \\ 
                        &  & & & HE & 0.954 & 0.038 & 807 \\ 
     
    \hline
\end{tabular}
\end{center}
\label{tab:designs}
\end{table*}
Thus, these two designs were considered in the second simulation study across different scenarios for the odds ratio. 

\subsection{Evaluating the Final Trial Design}
\begin{figure*}[h]
\setlength{\fboxsep}{0pt}%
\setlength{\fboxrule}{0pt}%
\begin{center}
\includegraphics[width =\textwidth]{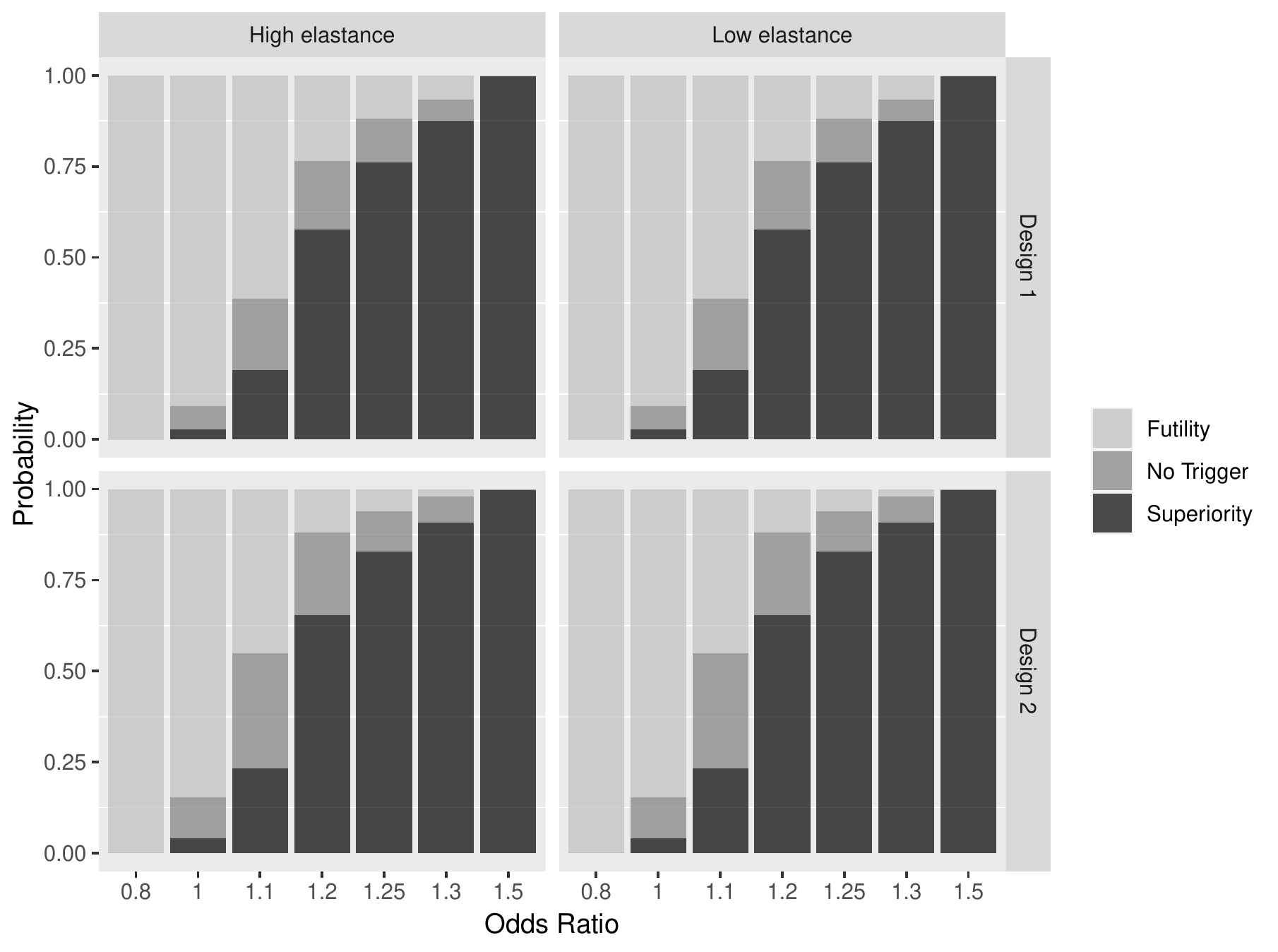}
\end{center}
\caption{For each odds ratio considered in our simulations, the probability of concluding futility (light grey), superiority (black) and reaching no statistical trigger (mid grey) by the maximum potential recruitment of 2,000 patients per state for the DRIVE domain is displayed. The results are separated by the High and Low elastance states. Results are separated for the two chosen trial designs, characterised by an initial recruitment level $N_{init}$ of 500 and 200, a threshold for declaring superiority $p_{sup}$ of 0.9925 and 0.995 and a threshold for declaring futility $p_{futi}$ of 0.95 and 0.95, respectively for Design 1 (top panels) and Design 2 (bottom panels). \label{barplots}}
\end{figure*}

The results of the second simulation phase are presented in Table \ref{tab:results} and Figure \ref{barplots}. From Figure \ref{barplots}, we can see that chance of reaching a statistical trigger was similar for both designs but the Design 1 had a higher probability of concluding superiority of DPL compared to Design 2. Thus, when DPL was superior to UC, Design 2 had a higher chance of incorrectly concluding futility. 

Table \ref{tab:results} provides the numerical results for the scenarios where the odds ratios are the same across the two severity states. The results for the differential treatment effects are presented in the supplementary material. Two designs offered similar type 1 errors. For most of the cases, Design 2 reduced the sample size slightly compared the Design 1 as both the 50\textsuperscript{th} and 80\textsuperscript{th} percentiles for the sample size were smaller. However, Design 1 had higher probability of concluding superiority when DPL is indeed superior compared to Design 2. The 80\textsuperscript{th} percentiles for the sample size exceed 2000 when the intervention was effective but the odds ratio was below or equal to the threshold chosen to represent futility, i.e., the odds ratio equals 1.1 and 1.2. 

\begin{table*}[!h]
\caption{Probability of achieving the statistical trigger of superiority and futility by the maximum potential recruitment of 2,000 patients per state for the DRIVE domain and the 50\% and 80\% percentiles for the sample size in lower and higher respiratory system elastance group. Results are displayed for the two chosen trial designs, characterised by an initial recruitment level $N_{init}$ of 500 and 200, a threshold for declaring superiority $p_{sup}$ of 0.9925 and 0.995 and a threshold for declaring futility $p_{futi}$ of 0.95 and 0.95, respectively for Design 1 and Design 2.}
\begin{center}
\begin{tabular}{|c|c|c|c|c|c|c|}
    \hline
    Odds & \multirow{2}{*}{State} &\multicolumn{3}{c}{Probability of}&
    \multicolumn{2}{|c|}{\makecell{Sample Size \\ Percentile}}\\
    \cline{3-7}
    Ratio && Superiority & Futility & No Trigger & 50th & {80th}\\
    \hline
    \multicolumn{7}{|c|}{Design 1} \\
    \hline
    \multirow{2}{*}{0.8}& LE & 0.000 & 0.999 & 0.001 & 534 & 625 \\ 
                        & HE & 0.000 & 1.000 & 0.000 & 522 & 558 \\  
    \hline
    \multirow{2}{*}{1}  & LE & 0.033 & 0.849 & 0.118 & 775 & 1525 \\ 
                        & HE & 0.035 & 0.856 & 0.109 & 665 & 1408 \\ 
    \hline
    \multirow{2}{*}{1.1} & LE & 0.230 & 0.458 & 0.312 & 1276 & 2002 \\ 
                         & HE & 0.246 & 0.502 & 0.252 & 1150 & 2001 \\ 
    \hline
    \multirow{2}{*}{1.2} & LE & 0.629 & 0.124 & 0.247 & 1132 & 2001 \\ 
                         & HE & 0.646 & 0.134 & 0.220 & 1073 & 2011 \\ 
    \hline
    \multirow{2}{*}{1.25} & LE & 0.801 & 0.051 & 0.148 & 899 & 1743 \\ 
                          & HE & 0.818 & 0.052 & 0.130 & 855 & 1633 \\
    \hline
    \multirow{2}{*}{1.3} & LE & 0.907 & 0.023 & 0.070 & 754 & 1386 \\ 
                         & HE & 0.919 & 0.032 & 0.049 & 692 & 1214 \\ 
    \hline
    \multirow{2}{*}{1.5} & LE & 1.000 & 0.000 & 0.000 & 609 & 639 \\ 
                         & HE & 1.000 & 0.001 & 0.001 & 533 & 594 \\ 
    \hline
    \multicolumn{7}{|c|}{Design 2} \\
    \hline
    \multirow{2}{*}{0.8}& LE & 0.000 & 1.000 & 0.000 & 258 & 388 \\ 
                        & HE & 0.001 & 0.999 & 0.000 & 256 & 320 \\ 
    \hline
    \multirow{2}{*}{1}   & LE & 0.033 & 0.849 & 0.118 & 775 & 1525 \\ 
                         & HE & 0.035 & 0.856 & 0.109 & 665 & 1408 \\ 
    \hline
    \multirow{2}{*}{1.1} & LE & 0.202 & 0.500 & 0.298 & 1079 & 2002 \\ 
                         & HE & 0.189 & 0.545 & 0.266 & 979 & 2001 \\ 
    \hline
    \multirow{2}{*}{1.2} & LE & 0.585 & 0.160 & 0.255 & 1035 & 2007 \\ 
                         & HE & 0.595 & 0.174 & 0.231 & 966 & 2007 \\
    \hline
    \multirow{2}{*}{1.25} & LE & 0.758 & 0.097 & 0.145 & 879 & 1702 \\ 
                          & HE & 0.770 & 0.097 & 0.133 & 774 & 1652 \\ 
    \hline
    \multirow{2}{*}{1.3} & LE & 0.876 & 0.060 & 0.064 & 734 & 1393 \\ 
                         & HE & 0.887 & 0.045 & 0.068 & 638 & 1284 \\ 
    \hline
    \multirow{2}{*}{1.5} & LE & 0.994 & 0.005 & 0.001 & 380 & 640 \\ 
                         & HE & 0.997 & 0.003 & 0.000 & 313 & 545 \\ 
    \hline
\end{tabular}
\end{center}
\label{tab:results}
\end{table*}

Table \ref{tab:results} confirms that Design 1 offers higher power, but also requires a slightly larger sample size, with the median sample size increased by between 20 and 176 for Design 1, compared to Design 2. The difference in sample size was more prominent when the probability of reaching either statistical trigger was high, especially for when the odds ratios are 0.8 and 1.5 across two states.

\begin{figure*}[!h]
\centering
\includegraphics[width =0.8\textwidth]{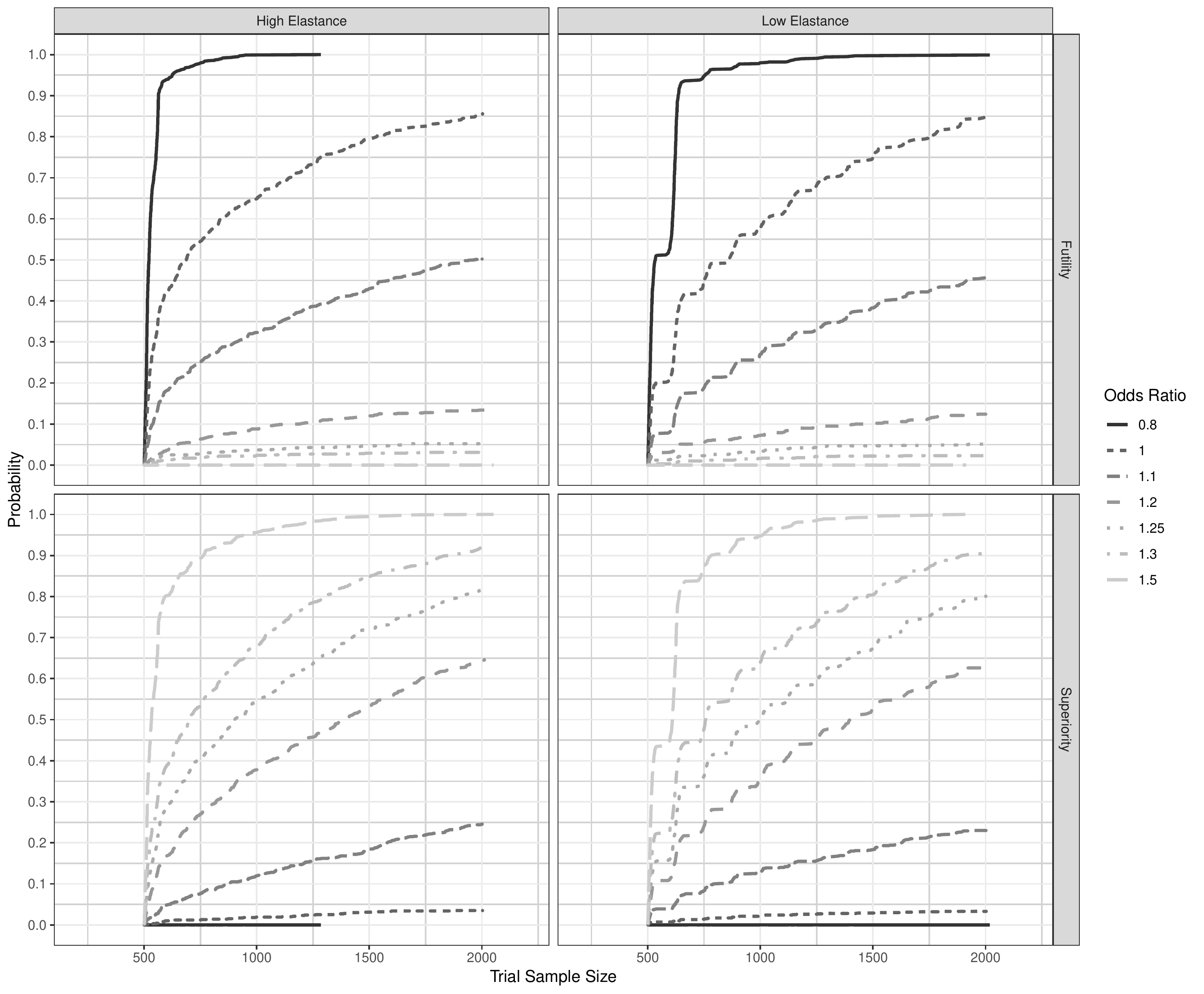}
\caption{\label{results} The cumulative probability of concluding futility, superiority and not reaching any statistical trigger for the novel driving pressure limited (DPL) intervention plotted against the number of patients recruited for design 1. Each line represented a different value (0.8, 1, 1.1, 1.2, 1.25, 1.3, 1.5) for the true odds ratio (OR) and the results are plotted separately for the two elastance states.}
\end{figure*}

Based on these results, Design 1 was chosen as the study design for DRIVE since it provides higher power without a substantial increase in sample size for more realistic scenarios. Finally, Figure \ref{results} represents the cumulative probability of declaring futility, superiority and not reaching either statistical trigger as the sample size increases for Design 1. Each odds ratio is plotted as a line and the results are provided separately for the high and low elastance groups. For scenarios associated with small and large odds ratios (0.8 and 1.5), the trial conclusions were reached quickly and the risk of incorrect conclusions (superiority for 0.8 and 1, futility for 1.25, 1.3 and 1.5) were small. For odds ratios closer to 1.2, the threshold for declaring futility, trial conclusions were reached at larger sample sizes and there was a steadier decline in the proportion of scenarios that had not reached a statistical trigger. This is because the evidence for superiority or futility will inevitably be weaker.

\subsection{Computational Efficiency of INLA}
The posterior distribution for the model parameters was computed approximately 620,000 times in this study. Estimating the posterior distributions using \texttt{stan}\cite{carpenter2017stan} within the \texttt{brms} package\cite{burkner2017brms} in \texttt{R} required around 8 minutes with approximately 500 patients across the two states and 1 hour and 37 minutes with approximately 4000 patients across the two states. This compares to 2.5 seconds and 13.1 seconds, respectively with INLA, implemented through the R-INLA package.\cite{martins2013bayesian} Thus, while the total CPU time with INLA was approximately 20 days, the required computation time with \texttt{stan} through \texttt{brms} would have been over 9 years of CPU time. While the efficiency of \texttt{stan} could have been improved using pre-complied models, a bespoke likelihood and alternative efficient implementations, it is unlikely that our simulation strategy could have been implemented in \texttt{stan}.

\section{Discussion}
The DRIVE trial aims to evaluate interventions for providing respiratory support in the ICU using an ordinal outcome that combines reductions in mortality and the days spent receiving ventilatory support. We developed a Bayesian adaptive trial design for DRIVE that stops for either superiority or futility of the novel intervention. Frequent analyses will allow the DRIVE trial to stop as soon as sufficient evidence is available to form conclusions about the intervention effect, or lack thereof.

Computationally intensive simulations were required to determine the appropriate thresholds to stop the DRIVE trial and the number of patients that should be recruited before starting the analyses. The ordinal outcome prohibited the use of conjugate models and would have restricted the number of scenarios that could have been considered if MCMC methods were required to determine the relevant posterior distributions. However, INLA allowed us to comprehensively assess 64 design scenarios and 22 treatment effect scenarios for the two selected designs. Thus, we determined an efficient design for the DRIVE trial and confirmed the validity of proposed analyses.

INLA can fit sufficiently wide range of models that most Bayesian adaptive trials could be efficiently evaluated. INLA easily allows for hierarchical modeling alongside a range of flexible linear and non-linear regression models. One limitation for the DRIVE trial was the restricted number of ordinal outcome categories that could be included while maintaining the efficiency of the INLA algorithm. This may lead to minimal differences in the performance for the proposed analysis on the 30 level outcome. However, despite this limitation, the efficiency of INLA allowed for a substantially more comprehensive evaluation of the DRIVE design within a realistic timeframe than using MCMC methods. 

\section{Conclusion}
We used the INLA algorithm to develop and evaluate a Bayesian adaptive trial that will compare novel mechanical ventilation strategies for hospitalised patients with AHRF. The computational efficiency of INLA allowed a comprehensive evaluation of this design to determine the optimal design, accounting for potential treatment effect heterogeneity.

\subsection*{Acknowledgements}
We would like to acknowledge the Statistical Analysis Committee for the PRACTICAL platform trial and all collaborators on the DRIVE trial, including our patient partners.

\subsection*{Funding}
\textbf{The author(s) disclosed receipt of the following financial support for the research, authorship, and/or publication of this article}: Anna Heath is supported by a Canada Research Chair in Statistical Trial Design and funded by the Discovery Grant Program of the Natural Sciences and Engineering Research Council of Canada (RGPIN-2021-03366).

\subsection*{Declaration of Conflicting Interests} None declared.

\bibliographystyle{sagev}
\bibliography{BIB}

\clearpage

\section*{Supplementary Material}

\onecolumn

\begin{table*}[ht]
\caption{The lower quartile, median, mean and upper quartile for the ventilator free days outcome for each of the odds ratios considered in our simulation study, reported separately for the two elastance states. The probability of death is also reported.}
\begin{center}
\begin{tabular}{cccccccc}
\hline
Elastance & Odds & \multicolumn{4}{c}{Ventilator Free Days} & Probability \\ \cline{3-6}
State & Ratio & Lower Quartile & Median & Mean & Upper Quartile & of Death \\
\hline
    Low & 0.8 & -1 & 16 & 13.1 & 25 & 0.33 \\
    Low & 1 & -1 & 19 & 14.6 & 26 & 0.28 \\
    Low & 1.1 & -1 & 20 & 15.2 & 26 & 0.26 \\
    Low & 1.2 & 0 & 20 & 15.7 & 26 & 0.24 \\
    Low & 1.25 & 0 & 21 & 16.0 & 26 & 0.23 \\
    Low & 1.3 & 0 & 21 & 16.2 & 26 & 0.23 \\
    Low & 1.5 & 5.5 & 22 & 17.1 & 27 & 0.2 \\
    \hline
    High & 0.8 & -1 & 0 & 9.1 & 21.75 & 0.45\\
    High & 1 & -1 & 6.5 & 10.5 & 23 & 0.4 \\
    High & 1.1 & -1 & 12 & 11.1 & 24 & 0.38\\
    High & 1.2 & -1 & 12.5 & 11.8 & 26 & 0.36 \\
    High & 1.25 & -1 & 13 & 12.0 & 24 & 0.35\\
    High & 1.3 & -1 & 14 & 12.3 & 25 & 0.34\\
    High & 1.5 & -1 & 16 & 13.2 & 25 & 0.31\\
    \hline
\end{tabular}
\end{center}
\label{tab:quartiles}
\end{table*}

\begin{table*}[ht]
\caption{Probability of achieving the statistical trigger of superiority and futility and of no reaching a statistical trigger at the maximum feasible sample size, rounded to 3 decimal places. The 50\% and 80\% percentiles for the sample size of the adaptive trial. Results are split by the low and high elastance states, LE and HE respectively and are displayed for the design with an initial recruitment level $N_{init}$ of 500, a threshold for declaring superiority $p_{sup}$ of 0.9925 and a threshold for declaring futility $p_{futi}$ of 0.95.}
\begin{center}
\begin{tabular}{|c|c|c|c|c|c|c|}
    \hline
    Odds & \multirow{2}{*}{State} &\multicolumn{3}{c}{Probability of}&
    \multicolumn{2}{|c|}{\makecell{Sample Size \\ Percentile}}\\
    \cline{3-7}
    Ratio && Superiority & Futility & No Trigger & 50th & {80th}\\
    \hline
    0.8 & LE & 0.001 & 0.999 & 0.000 & 599 & 627 \\ 
    1.0 & HE & 0.030 & 0.915 & 0.055 & 563 & 1111 \\ 
    \hline
    1.1 & LE & 0.213 & 0.493 & 0.294 & 1247 & 2001 \\ 
    1.0  & HE & 0.037 & 0.853 & 0.110 & 744 & 1506 \\ 
    \hline
    1.2 & LE & 0.574 & 0.180 & 0.246 & 1135 & 2002 \\ 
    1.0 & HE & 0.057 & 0.802 & 0.141 & 795 & 1606 \\ 
    \hline
    1.25 & LE & 0.743 & 0.082 & 0.175 & 1012 & 1883 \\ 
    1.00 & HE & 0.069 & 0.803 & 0.128 & 819 & 1559 \\
    \hline
    1.3 & LE & 0.883 & 0.044 & 0.073 & 877 & 1507 \\ 
    1.0 & HE & 0.080 & 0.793 & 0.127 & 790 & 1559 \\
    \hline
    1.5 & LE & 0.999 & 0.001 & 0.000 & 620 & 868 \\ 
    1.0 & HE & 0.092 & 0.793 & 0.115 & 829 & 1558 \\ 
    \hline
    1.3 & LE & 0.899 & 0.024 & 0.077 & 761 & 1430 \\ 
    1.2 & HE & 0.683 & 0.114 & 0.203 & 969 & 2002 \\ 
    \hline
    1.3 & LE & 0.888 & 0.030 & 0.082 & 866 & 1500 \\ 
    1.1 & HE & 0.324 & 0.393 & 0.282 & 1153 & 2007 \\
    \hline \hline
    1.0 & LE & 0.033 & 0.876 & 0.091 & 637 & 1269 \\ 
    0.8 & HE & 0.000 & 1.000 & 0.000 & 524 & 561 \\ 
    \hline
     1.0 & LE & 0.047 & 0.816 & 0.137 & 865 & 1673 \\ 
     1.1 & HE & 0.209 & 0.532 & 0.259 & 1056 & 2007 \\ 
    \hline
    1.0 & LE & 0.059 & 0.794 & 0.147 & 896 & 1764 \\ 
    1.2 & HE & 0.546 & 0.210 & 0.244 & 1171 & 2005 \\ 
    \hline
    1.0 & LE & 0.063 & 0.784 & 0.153 & 900 & 1748 \\ 
    1.25 & HE & 0.750 & 0.110 & 0.140 & 1002 & 1693 \\  
    \hline
    1.0 & LE & 0.062 & 0.779 & 0.159 & 986 & 1759 \\ 
    1.3 & HE & 0.859 & 0.062 & 0.079 & 845 & 1412 \\ 
    \hline
    1.0 & LE & 0.104 & 0.757 & 0.139 & 1001 & 1652 \\ 
    1.5 & HE & 0.998 & 0.002 & 0.000 & 562 & 751 \\ 
    \hline
    1.2 & LE & 0.666 & 0.109 & 0.225 & 1109 & 2003 \\ 
    1.3 & HE & 0.901 & 0.030 & 0.069 & 751 & 1340 \\ 
    \hline
    1.1 & LE & 0.298 & 0.401 & 0.301 & 1243 & 2001 \\ 
    1.3 & HE & 0.874 & 0.049 & 0.077 & 825 & 1394 \\ 
    \hline 
\end{tabular}
\end{center}
\label{tab:diffresults500}
\end{table*}

\begin{table*}[ht]
\caption{Probability of achieving the statistical trigger of superiority and futility and of no reaching a statistical trigger at the maximum feasible sample size, rounded to 3 decimal places. The 50\% and 80\% percentiles for the sample size of the adaptive trial. Results are split by the low and high elastance states, LE and HE respectively and are displayed for the design with an initial recruitment level $N_{init}$ of 200, a threshold for declaring superiority $p_{sup}$ of 0.995 and a threshold for declaring futility $p_{futi}$ of 0.95.}
\begin{center}
\begin{tabular}{|c|c|c|c|c|c|c|}
    \hline
    Odds & \multirow{2}{*}{State} &\multicolumn{3}{c}{Probability of}&
    \multicolumn{2}{|c|}{\makecell{Sample Size \\ Percentile}}\\
    \cline{3-7}
    Ratio && Superiority & Futility & No Trigger & 50th & {80th}\\
    \hline
    0.8 & LE & 0.001 & 0.999 & 0.000 & 262 & 501 \\ 
    1.0 & HE & 0.019 & 0.928 & 0.055 & 331 & 894 \\ 
    \hline
    1.1 & LE & 0.195 & 0.537 & 0.268 & 998 & 2009 \\ 
    1.0 & HE & 0.053 & 0.853 & 0.094 & 622 & 1430 \\ 
    \hline
    1.2 & LE & 0.488 & 0.247 & 0.265 & 1123 & 2002 \\ 
    1.0 & HE & 0.062 & 0.822 & 0.116 & 650 & 1530 \\ 
    \hline
    1.25 & LE & 0.647 & 0.147 & 0.206 & 998 & 2001 \\ 
    1.0 & HE & 0.066 & 0.789 & 0.145 & 715 & 1651 \\ 
    \hline
    1.3 & LE & 0.783 & 0.098 & 0.119 & 858 & 1636 \\ 
    1.0 & HE & 0.088 & 0.777 & 0.135 & 736 & 1665 \\
    \hline
    1.5 & LE & 0.984 & 0.015 & 0.001 & 505 & 878 \\ 
    1.0 & HE & 0.146 & 0.751 & 0.103 & 704 & 1489 \\  
    \hline
    1.3 & LE & 0.852 & 0.066 & 0.082 & 757 & 1480 \\ 
    1.2 & HE & 0.649 & 0.145 & 0.206 & 929 & 2026 \\ 
    \hline
     1.3 & LE & 0.808 & 0.085 & 0.107 & 761 & 1520 \\ 
     1.1 & HE & 0.332 & 0.401 & 0.267 & 984 & 2021 \\ 
    \hline \hline
    1.0 & LE & 0.018 & 0.891 & 0.091 & 495 & 1239 \\ 
    0.8 & HE & 0.000 & 1.000 & 0.000 & 262 & 424 \\ 
    \hline
    1.0 & LE & 0.048 & 0.831 & 0.121 & 645 & 1624 \\ 
    1.1 & HE & 0.158 & 0.606 & 0.236 & 849 & 2001 \\  
    \hline
    1.0 & LE & 0.056 & 0.803 & 0.141 & 755 & 1667 \\ 
    1.2 & HE & 0.450 & 0.293 & 0.257 & 1087 & 2032 \\  
    \hline
    1.0 & LE & 0.061 & 0.786 & 0.153 & 775 & 1673 \\ 
    1.25 & HE & 0.646 & 0.205 & 0.149 & 925 & 1851 \\   
    \hline
     1.0 & LE & 0.070 & 0.780 & 0.150 & 782 & 1739 \\ 
     1.3 & HE & 0.758 & 0.135 & 0.107 & 824 & 1589 \\
    \hline
    1.0 & LE & 0.120 & 0.729 & 0.151 & 865 & 1733 \\ 
    1.5 & HE & 0.977 & 0.022 & 0.001 & 503 & 824 \\  
    \hline
    1.2 & LE & 0.618 & 0.155 & 0.227 & 1002 & 2001 \\ 
    1.3 & HE & 0.869 & 0.055 & 0.076 & 689 & 1381 \\ 
    \hline
     1.1 & LE & 0.277 & 0.421 & 0.302 & 1153 & 2004 \\ 
     1.3 & HE & 0.807 & 0.095 & 0.098 & 772 & 1515 \\ 
    \hline
\end{tabular}
\end{center}
\label{tab:diffresults350}
\end{table*}
\clearpage 

\section*{Dynamic Borrowing}

\begin{figure}

  \centering
  \includegraphics[width=0.8\linewidth]{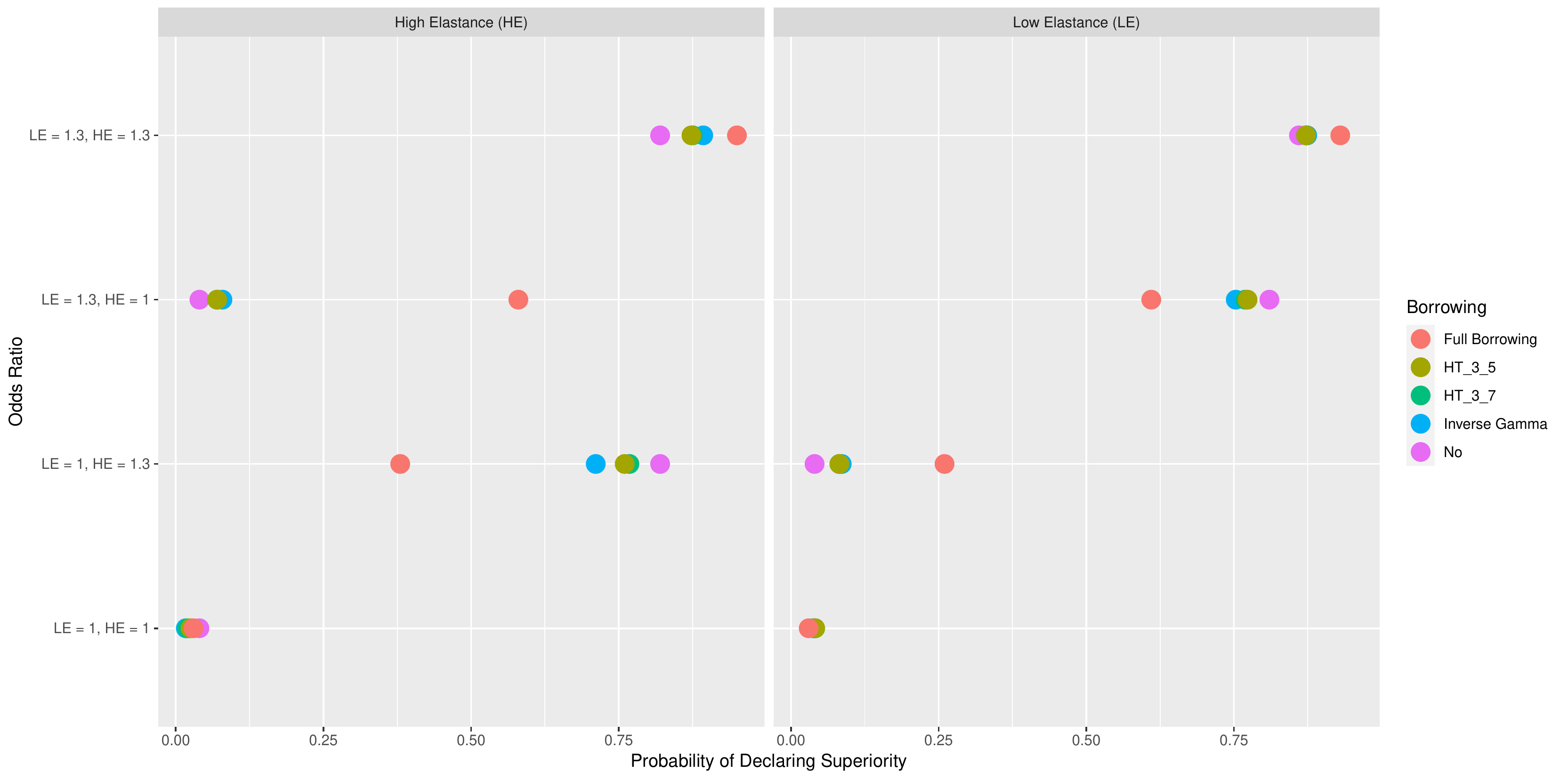}
  \caption{Probability of Concluding Superiority Under Different Odds Ratio Setting}
  \label{fig:sfig1}

  \centering
  \includegraphics[width=0.8\linewidth]{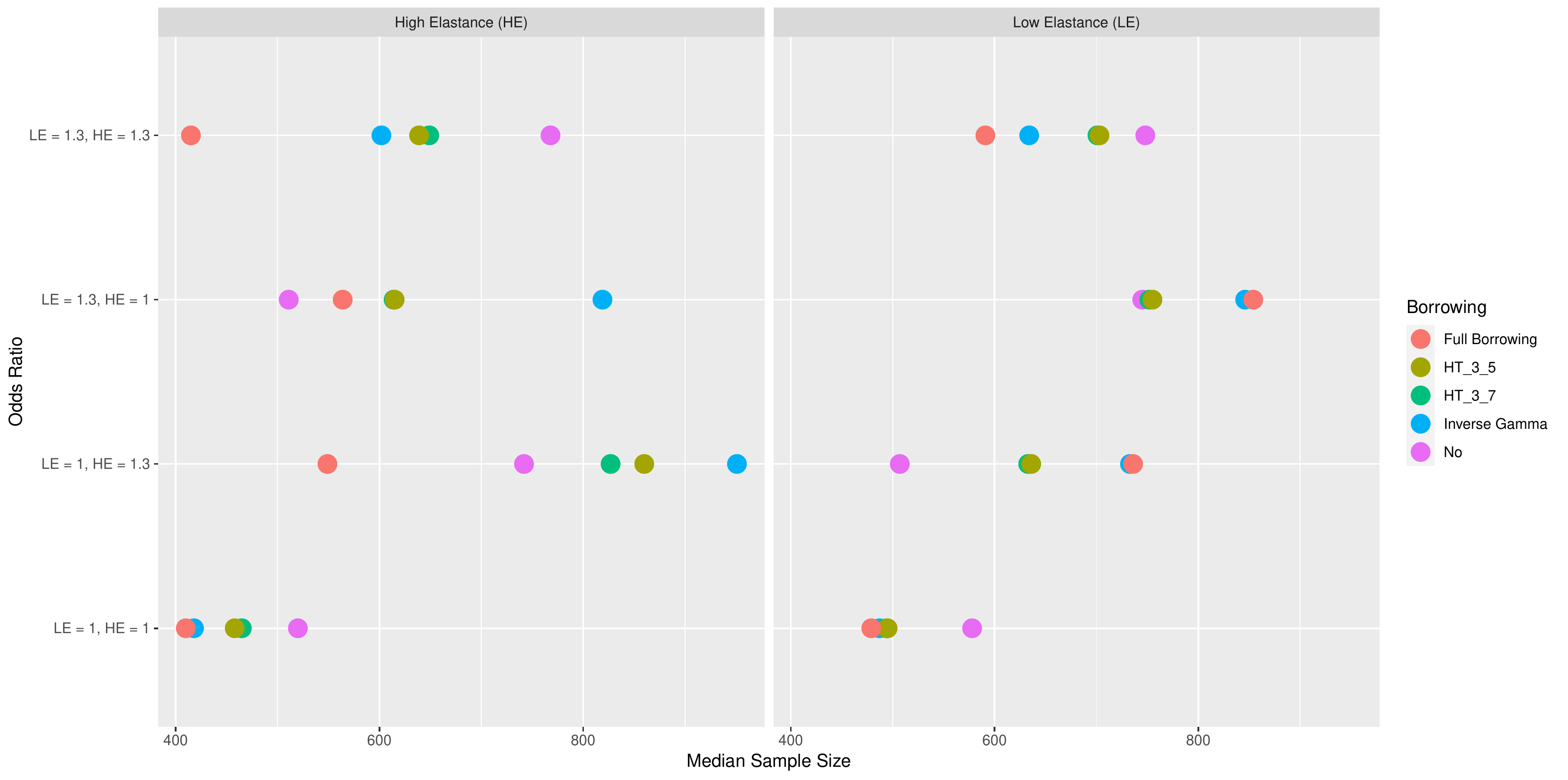}
  \caption{$50^{th}$ Percentile of Sample Size Under Different Odds Ratio Setting}
  \label{fig:sfig2}

  \centering
  \includegraphics[width=0.8\linewidth]{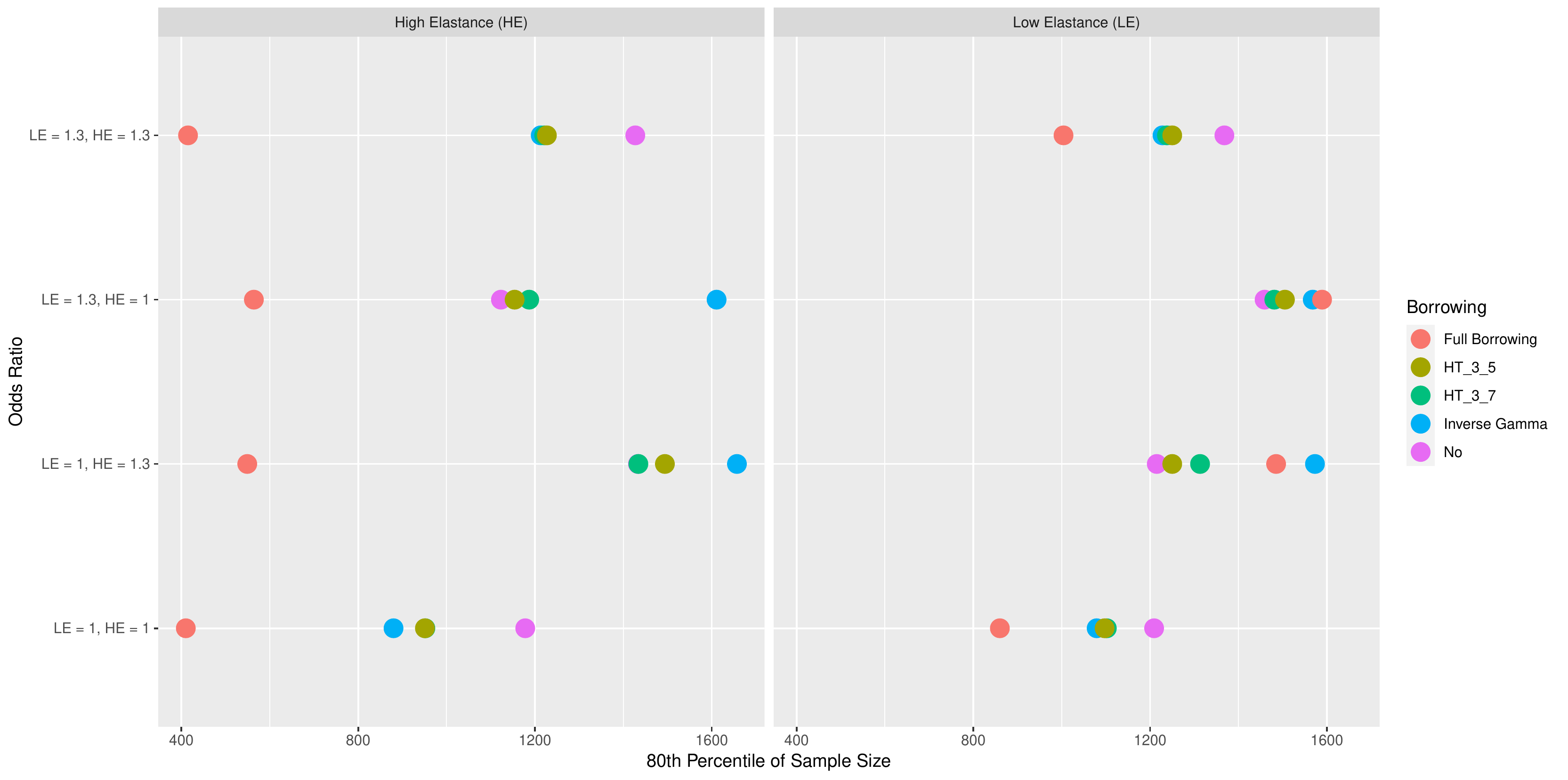}
  \caption{$80^{th}$ Percentile of Sample Size Under Different Odds Ratio Setting}
  \label{fig:sfig3}
\caption{The effect of dynamic borrowing using different priors on (A) probability of concluding superiority, (B) $50^{th}$ percentile of sample size and (C) $80^{th}$ percentile of sample size under different treatment effect settings}
\label{Figure DB}
\end{figure}

The proportional odds model fit in the DRIVE trial design uses dynamic borrowing to account for differences in the treatment effect across severity states, while borrowing strength when the treatment effects are similar. When choosing the prior to incorporate dynamic borrowing to account for information from different severity states, we evaluated three potential priors listed below:
\begin{enumerate}
    \item Inverse Gamma prior with shape parameter $\alpha = 0.125$ and scale parameter $\beta = 355.56$.
    \item Half-t prior with 3 degrees of freedom and a scale parameter of 5.
    \item Half-t prior with 3 degrees of freedom and a scale parameter of 7.
\end{enumerate}
We then plotted the effect of adopting dynamic borrowing in the modeling for all these priors, and compared them with the scenarios where there is no borrowing at all and full borrowing, i.e., a single treatment effect estimated across both states. The results are displayed in Figure \ref{Figure DB}.\\
Sub-Figure \ref{Figure DB}-A shows the probability of declaring superiority, Sub-Figure \ref{Figure DB}-B shows the $50^{th}$ percentile of trial sample sizes and Sub-Figure \ref{Figure DB}-C shows the $80^{th}$ percentile of trial sample sizes. Each plot is separated by the two severity states across a range of treatment effects; both severity state have the same odds ratio (either 1 or 1.3) and the two crossed scenarios where one severity state has odds ratio 1 and the other odds ratio 1.3. 

When not considering dynamic borrowing, the full borrowing leads to a substantial increase in the type 1 error rates when the treatment effects are different. The no borrowing scenario results in much lower power and much bigger sample size when the treatment effects are the same. Adopting dynamic borrowing makes the trial design more efficient by resulting in smaller type 1 error rates, smaller sample size, and providing additional power when treatment effects are the same.

Out of all priors that were evaluated, we chose the half-t prior with 3 degrees of freedom and a scale parameter of 7 over the other two options. It provides similar power but results in much smaller sample size compared to the other half-t prior and the inverse Gamma prior when appropriate.

\end{document}